\documentclass[11pt,a4paper]{article}
    \usepackage{jheppub}
    \usepackage[T1]{fontenc}     
   \usepackage{subcaption}
   \usepackage{float}
   \usepackage{array}

\DeclareMathOperator{\Tr}{Tr}

\title{Classifying large N limits of multiscalar theories by algebra}

\author{Nadia Flodgren and Bo Sundborg}

\affiliation{The Oscar Klein Centre \& Department of Physics, Stockholm University,\\
AlbaNova, 106 91 Stockholm, Sweden}

\emailAdd{nadia.flodgren@fysik.su.se}
\emailAdd{bo.sundborg@fysik.su.se}

\abstract{We develop a new approach to RG flows and show that one-loop flows in multiscalar theories can be described by commutative but non-associative algebras. As an example related to $D$-brane field theories and tensor models, we study the algebra of a theory with $M$ $SU(N)$ adjoint scalars and its large $N$ limits. The algebraic concepts of idempotents and Peirce numbers/Kowalevski exponents are used to characterise the RG flows. 
We classify and describe all large N limits of algebras of multiadjoint scalar models: the standard `t Hooft matrix theory limit, a `multi-matrix' limit, each with one free parameter, and an intermediate case with extra symmetry and no free parameter of the algebra, but an emergent free parameter from a line of one-loop fixed points. The algebra identifies these limits without diagrammatic or combinatorial analysis.}

\begin{document}

\maketitle

\section{Introduction}

Recently, purely scalar field theories have been proposed to be well-defined in the ultraviolet by unconventional choices of RG (renormalisation group) trajectories for the couplings. While the RG equations describe the flow of couplings as a function of scale, they naturally extend to complex couplings and do not themselves prescribe signs or reality properties. Examples of unconventional solutions include negative couplings in the far UV, related to PT symmetric (rather than Hermitean) quantum field theories \cite{Romatschke2022,Romatschke2023}, and trajectories where some couplings are purely imaginary \cite{BenedettiGurauHarribey2019,Benedetti:2019ikb,Benedetti:2020sye,BergesGurauPreis2023}, which nonetheless have been argued to yield quantum field theories with real-valued observables. In both cases, the theories appear to be asymptotically free, a property which conventionally is reserved for non-Abelian gauge theories in 4d. These developments prompt us to reconsider the weak-coupling behaviour of purely scalar field theories. In addition to the standard interest of weak coupling results for typically IR free scalar field theories, weak scalar couplings can thus become important in the UV\footnote{In contrast to the massless IR theories, massless UV theories do not require fine-tuning.}.  Certain large $N$ limits in which the effects of gauge interactions are suppressed \cite{Ferrari:2017ryl} also motivate the study of purely scalar theories.

In the mathematical literature, a connection between non-associative algebras and dynamical systems  similar to the RG equations has been known from the work of Markus \cite{MR0132743}, and the interest in this approach has recently been revived \cite{MR4191467,math11081790}. As we will show, some of these new developments are particularly useful in the class of dynamical systems which arise as RG equations of 4d quantum field theory.

In a renormalisable quantum field theory a finite number of coupling constants typically become scale dependent, and part of the perturbation expansion characterises this scale dependence in terms of beta functions, which are functions of the coupling constants and determine the flow vector field of the RG flow, ie the rate of change of the couplings with respect to the scale. The RG equations are simply the system of first order ordinary differential equations defined by the beta function vector field on the space of couplings. While the beta functions can be calculated perturbatively from Feynman diagrams, as the RG equations represent perturbations in several parameters, the solutions are still potentially quite intricate \cite{Bosschaert:2021ooy}.

In four dimensions, precise methods to probe interacting quantum field theories are scarce except in the presence of supersymmetry, and sorely needed. Large $N$ methods which exploit simplifications in theories with a large number of fields, $N$, are therefore especially important in non-supersymmetric theories. In particular, large $N$ arguments have established conformal and near-conformal field theories in four dimensions by extending the scope of perturbation theory, in weakly interacting gauge theories \cite{Caswell:1974gg,Banks:1981nn,Litim:2014uca}. The present work develops this framework by studying  alternative large $N$ limits by algebraic methods.

In this paper we study an algebraic characterisation of the lowest order RG equations, which we rediscovered in previous work \cite{Flodgren:2023lyl}. It had in fact been studied before in $4 - \epsilon$ dimensions by Michel \cite{Michel:1983in}, but we believe it to be especially significant in 4 dimensions. Michel used the algebra to search for fixed points of the RG flow describing phase transitions at thermal equilibrium in 3 dimensions. Assuming regularity of the fixed points and the $\epsilon$ expansion, he related the fixed points to idempotents of the algebra (explained in our section \ref{sec_3}) non-perturbatively. In contrast, our use of the algebra in 4 dimensions is manifestly perturbative. This is dictated by its definition (in terms of the one loop beta function) and that the couplings in the algebra are marginal at tree level in d= 4. Our focus on the purely perturbative aspects of the algebra might seem restrictive, but we believe perturbation theory to be the proper place of the algebra --- where it will be most illuminating --- especially when combined with large $N$ limits. The present work begins to uncover the potential of the algebraic perspective in perturbation theory.

An interesting feature of the algebraic formulation is that it combines several distinct Feynman diagrams into one algebraic object.  
An algebraic structure that can rein in the rapidly increasing number of Feynman diagrams would be immensely useful at higher orders. 
For example, since large $N$ limits of vectorial type, like the $O(N)$ model, are one-loop exact, the one-loop algebra completely characterises the leading behaviour in such limits \cite{Jepsen:2023pzm}. 
While a more general higher order picture is the ultimate motivation for our detailed exploration of the algebraic structure, it is beyond the scope of the present study. Instead, we find a more immediate reward: a general understanding of the RG flow patterns of weakly coupled scalar field theories, with examples of considerable simplifications in large $N$ limits respecting the algebras.

We focus our study on large $N$ limits of 4d theories with $M$ adjoint  $SU(N)$ multiplets and unbroken $SU(N)\times O(M)$ symmetry, but it will become apparent that the algebraic framework we unravel is independent of these details, and only the specifics of the algebras depend on the theory. Our choice of example theories is motivated by relations to 3D-brane theories \cite{Flodgren:2023lyl} and various interesting large $N$ theories \cite{Ferrari:2017ryl,Azeyanagi:2017mre,Carrozza:2020eaz} related to tensor field theories \cite{Klebanov:2017nlk,Benedetti:2020sye}. Indeed, by simple scaling arguments we find directly the same large $N$ limits of such theories, which were identified before by advanced combinatorial arguments.

This paper is organized in the following way. Section \ref{sec_2} reviews the fundamentals of the non-associative algebraic description of one-loop RG equations. This description is then applied to a multiscalar model with $SU(N)\times O(M)$ symmetry, which is studied for three distinct large $N$ limits. In section \ref{sec_3} we characterise the algebra in detail using special elements of the algebra. The physical significance of these mathematical concepts is also discussed.
The one-loop RG flows for the large $N$ limits are shown in section \ref{sec_4}. In the conclusion we give an outlook on future directions.

\section{One-loop algebras and their large $N$ limits}\label{sec_2}

We begin by reviewing the algebra fundamentals we developed in \cite{Flodgren:2023lyl}. We go on to focus on a model with interesting large $N$ limits, which will serve as test cases for our methods. In a multiscalar model with $SU(N)\times O(M)$ symmetry we ask which scalings of the couplings, with respect to $N$ and $M$, lead to well-defined algebras and find three qualitatively different solutions. Lastly, we illuminate the structures of each limit algebra, in terms of subalgebras, ideals and quotient algebras.

\subsection{Algebra fundamentals}

In \cite{Flodgren:2023lyl} we developed a ``marginal algebra'' to describe the one-loop RG flow for a multiscalar gauge theory in four dimensions. 
Here we focus on the corresponding purely scalar theories, with no gauge bosons, but keeping massless scalars $\phi_{A}$ and their marginal quartic interactions in 4D. The Lagrangian is
\begin{equation}
\mathcal{L}_{\text{int}} = - \frac{1}{4!}\lambda_{ABCD}\phi_{A}\phi_{B}\phi_{C}\phi_{D}.
\end{equation}
The index $A$ is a symbolic multi-index\footnote{Note that in \cite{Flodgren:2023lyl} we denoted the multi-index by $\bar{A}$ instead of $A$.} which includes indices for the individual irreducible representations of the scalars and any global symmetries.
The quartic interaction tensor $\lambda_{ABCD}$ is chosen to be totally symmetric in the multi-indices. 
The one-loop beta functions for these theories are among the standard results found in for example \cite{Machacek:1984zw,Luo:2002ti}:
\begin{equation} \label{eq:beta}
\begin{split}
&\frac{d}{dt}{\lambda}_{ABCD}=\mu \frac{d}{d\mu}\lambda_{ABCD}=\beta_{ABCD} 
= \frac{1}{(4\pi)^2} \Lambda_{ABCD}^2\\
\end{split}
\end{equation}
where
\begin{equation} \label{eq_LLA}
\begin{split}
&\Lambda^2_{ABCD}=\frac{1}{8}\sum_{\text{perms}} \lambda_{ABEF}\lambda_{EFCD}.
\end{split}
\end{equation}
The scale $\mu$ runs from the IR to the UV, and $t=\log(\mu)$ is the ``time'' variable in the RG equations. The sum over the permutations refers to the permutations of the multi-indices. 

Let us review the reason for why an algebraic description of the one-loop RG flow for marginal couplings is possible. 
A renormalisable theory has a finite number of renormalised couplings $\lambda$, and of these we consider only the marginal ones. We can think about the marginal couplings as a vector space. 
The beta functions describe the RG flow of the couplings and are calculated up to a given loop-order using Feynman diagrams.  
For a purely scalar theory the one-loop order beta function \eqref{eq:beta} is quadratic in the couplings, $\beta_{\lambda}=\frac{1}{(4\pi)^2}P_2(\lambda)$ where $P_2(\lambda)$ is a quadratic polynomial in $\lambda$. 
This structure gives rise to a product of the couplings
\begin{equation} \label{eq_d0}
\begin{split}
\lambda \diamond \kappa &\equiv \frac{1}{2}(P_2(\lambda+\kappa)-P_2(\lambda)-P_2(\kappa)) \\
&=\frac{(4\pi)^2}{2}(\beta_{\lambda+\kappa}-\beta_{\lambda}-\beta_{\kappa}),
\end{split}
\end{equation}
since the right-hand side is arranged to only preserve the cross terms between $\lambda$ and $\kappa$. Because an algebra is a vector space with a distributive product, the vector space of the marginal couplings with the $\diamond$ product is an algebra. The $\diamond$ product is commutative but not by definition associative.

Let us next consider how to calculate the marginal algebra. 
First we introduce a complete basis of symmetric tensor structures\footnote{Note that in \cite{Flodgren:2023lyl} the basis elements were denoted by $g^k$, here they are denoted by $e^k$.} $e^k_{ABCD}$, with corresponding coupling constants $\lambda_k$ to describe the marginal quartic interaction 
\begin{equation} \label{eq_assume}
\begin{split}
\lambda_{ABCD} = \lambda_k e^k_{ABCD}.
\end{split}
\end{equation}
The basis is assumed to be closed under renormalisation group flow, ie it spans the space of marginal symmetric four-point couplings, labeled by $k$.  
The one-loop beta function $\beta_k=\frac{d\lambda_k}{dt}$ is in terms of this basis
\begin{equation} \label{eq_antz}
\begin{split}
\beta_{ABCD} = \beta_k e^k_{ABCD}.
\end{split}
\end{equation}
For a multiscalar theory $\beta_k$ has the form 
\begin{equation} \label{eq_bk1}
\begin{split}
\beta_k = \frac{1}{(4\pi)^2 }\lambda_m\lambda_n  C^{mn}_k.\\
\end{split}
\end{equation}
The coefficients $C^{mn}_k$ act as structure constants encoding how the couplings $\lambda_m$ and $\lambda_n$ appear in the beta function for $\lambda_k$. 
Their role as structure constants is evident in the expression for the $\diamond$ product. In terms of the basis, \eqref{eq_LLA} yields
\begin{equation} \label{eq_diamond}
\begin{split}
\left(e^m \diamond e^n\right)_{ABCD} \equiv \frac{1}{8}\sum_{\text{perms}} e^m_{ABEF}e^n_{EFCD}  
  = C^{mn}_ke^k_{ABCD}.
\end{split}
\end{equation}
The sum over permutations of the multi-indices guarantees that the product is totally symmetric. 
The symmetrisation in (\ref{eq_diamond}) is the reason for the non-associativity of the algebra.\footnote{An example of a generally non-associative product is provided by the commutative Jordan product $\circ$ of symmetric matrices $A$ and $B$, $A\circ B \equiv (AB+BA)/2$, which is a symmetrisation $(A_{ij}B_{jk}+A_{kj}B_{ji})/2=(A_{ij}B_{jk}+B_{ij}A_{jk})/2$.} 
Note that from now on we will mostly suppress all field indices and write $e^m \diamond e^n  = C^{mn}_k e^k$ as a shorthand for the product. 
The general definition of the $\diamond$ product (\ref{eq_d0}) and its representation (\ref{eq_diamond}) in terms of structure constants are guaranteed to be equal by the one-loop expressions for the beta functions. 

By \textit{calculating the algebra} we mean calculating all the structure constants $C^{mn}_k$ for a chosen basis. 
The algebra can also be represented as a symmetric multiplication table
\begin{equation}
\left(
\begin{array}{cccc}
e^{1}\diamond e^{1}  & \dots & e^{1}\diamond e^{K}\\
\vdots  & \ddots  & \vdots  \\
e^{K}\diamond e^{1}  &\dots & e^{K}\diamond e^{K}  
\end{array}
\right).
\end{equation}

Let us consider the $O(N)$ vector model in $d=4$ as an example.\footnote{Several other examples will be detailed and compared by N. Flodgren (work in progress).}  It has one quartic invariant $(\bar{\phi}^2)^2=(\phi_A\phi_A)^2$, where $A=1,\dots, N$, with the corresponding symmetric basis element
\begin{equation}
\begin{split}
e^1_{ABCD}= \delta_{AB}\delta_{CD}+\delta_{AC}\delta_{BD}+\delta_{AD}\delta_{BC} 
\end{split}
\end{equation}
and coupling constant $\lambda_1$. 
The explicit relationship between the basis element and the invariant is
\begin{equation}
\begin{split}
\frac{1}{4!}e^1_{ABCD}\phi_A\phi_B\phi_C\phi_D &= \frac{1}{8}(\bar{\phi}^2)^2. \\
\end{split}
\end{equation}
The algebra has a single structure constant
\begin{equation}
\begin{split}
C^{11}_1 = 8+N,
\end{split}
\end{equation}
which is easily seen from the 1-loop order beta function
\begin{equation}
\begin{split}
\beta_{\lambda}= \frac{1}{(4\pi)^2}(8+N)\lambda_1^2.
\end{split}
\end{equation}

\subsection{Adjoint multiscalar model}\label{sec_22}

Consider a model that has $M$ scalar multiplets in the adjoint representation of $SU(N)$ invariant under $O(M)$ symmetry and a global $SU(N)$ symmetry. The multi-indices are now decomposed into two indices $A=\bar{a}\bar{A}$, where $\bar{A}=1,\dots, N^2-1$ is the $SU(N)$-index and $\bar{a}=1,\dots,M$ is the scalar multiplet index. 
The scalar fields in terms of the multi-indices are 
\begin{equation}
\Phi_{\bar{a}}=\phi_{A}T_{\bar{A}}=\phi_{\bar{a}\bar{A}}T_{\bar{A}}, 
\end{equation}
where $T_{\bar{A}}$ are matrices in the fundamental representation normalised as $\Tr(T_{\bar{A}}T_{\bar{B}})=\frac{1}{2}\delta_{\bar{A}\bar{B}}$. The components $\phi_{\bar{a}\bar{A}}$ transform in ($M$ copies of) the adjoint representation of $SU(N)$. 

The interactions can be expressed as symmetric tensor structures or quartic invariant polynomials of the fields $\Phi_{\bar{a}}$. 
For this model there are four invariant polynomials needed to describe the symmetric potential: two single-trace (ST) and two double-trace (DT) polynomials. They correspond to the invariant tensor structures $e^{k}_{ABCD}$ via
\begin{equation} \label{eq_poly1}
\begin{aligned}
 \frac{1}{4 !} e_{ABCD}^{1 s} \phi_{A} \phi_{B} \phi_{C} \phi_{D}&=\frac{1}{2} \Tr \Phi_{\bar{a}} \Phi_{\bar{a}} \Phi_{\bar{b}} \Phi_{\bar{b}} \\
\frac{1}{4 !} e_{ABCD}^{1 t} \phi_{A} \phi_{B} \phi_{C} \phi_{D}&=\frac{1}{4} \Tr \Phi_{\bar{a}} \Phi_{\bar{b}} \Phi_{\bar{a}} \Phi_{\bar{b}} \\
\frac{1}{4 !} e_{ABCD}^{2 s} \phi_{A} \phi_{B} \phi_{C} \phi_{D}&=\frac{1}{2} \Tr \Phi_{\bar{a}} \Phi_{\bar{a}} \Tr \Phi_{\bar{b}} \Phi_{\bar{b}}  \\
 \frac{1}{4 !} e_{ABCD}^{2 t} \phi_{A} \phi_{B} \phi_{C} \phi_{D}&=\Tr \Phi_{\bar{a}} \Phi_{\bar{b}} \Tr \Phi_{\bar{a}} \Phi_{\bar{b}} .
\end{aligned}
\end{equation}
The expanded expressions for $e^{k}_{ABCD}$ are given by (\ref{eq_4g}) in Appendix \ref{App0}. Here the superscript $k=\{1s,1t,2s,2t\}$. The superscripts $1$ and $2$ stand for single-trace and double-trace respectively, while the $s$ and $t$ stand for products of $O(M)$ scalars or tensors, respectively. 
The invariant tensors structures $e^{k}_{ABCD}$ in (\ref{eq_poly1}) is our choice of basis for studying the algebra of this model. 
From (\ref{eq_poly1}) it is clear that these basis elements have different discrete symmetries. We will call such bases, with basis elements of definite symmetry properties, symmetry-respecting.

The algebra products of the basis elements (\ref{eq_poly1}) are given in Appendix \ref{App1} by equations (\ref{eq_Nprod1}), (\ref{eq_Nprod2}) and (\ref{eq_Nprod3}). Note the factors of $N$ and $M$ in the products. In order to have a well-defined algebra in the large $N$ and/or large $M$ limits it is necessary to rescale the basis elements.

\subsection{Rescaling the couplings}

The scalar couplings and their corresponding basis elements can be rescaled as $\lambda_k \rightarrow \Lambda_k$ and $e^k \rightarrow E^k$ respectively, provided they fulfil the relation
\begin{equation}\label{eq_lgLG}
\lambda_ke^k= \Lambda_k E^k.
\end{equation}
In order to study large $N$ theories we want algebras that directly characterise large $N$ theories. Thus, the goal with a rescaling is to obtain a finite multiplication matrix when we take the large $N$ limit. 

Let us start from our basis elements $e^k$, where $k=\{1s,1t,2s,2t\}$, and make the general rescaling
\begin{equation}\label{eq_eE}
e^k= N^{n(k)}M^{m(k)}E^k.
\end{equation}
$E^k$ are the rescaled elements with corresponding rescaled couplings $\Lambda_k$.  
The exponents $n(k)$ and $m(k)$ are non-negative real numbers. 
To take the simultaneous large $N$ and large $M$ limits we define a relationship between $N$ and $M$%. 
\begin{equation}\label{eq_MN}
M=v(a)N^a, 
\end{equation}
where $v(a)$ is a constant independent of $N$ and $M$, and $a\geq 0$ is a real number. 
The rescaling in terms of $N$ is then 
\begin{equation}\label{eq_MN2}
e^k= N^{n(k)+am(k)}v(a)^{m(k)}E^k.
\end{equation}

Next, we calculate the multiplication matrix for the basis elements $E^k$. The elements of the multiplication matrix have powers of $N$ containing 
\begin{equation}
    p(k)=n(k)+am(k),
\end{equation}
but not $n(k)$ and $m(k)$ individually. 
A well-defined algebra at large $N$ will have only non-divergent multiplication matrix elements in the large $N$ limit. Therefore, the exponents of $N$ in the multiplication matrix elements need to be less than or equal to zero. 
This requirement constrains $p(k)$ and indirectly $a$ (for a solution to exist).

For example, in the multiplication matrix for $E^k$ one of the terms in the product 
\begin{equation}
E^{1s} \diamond E^{1s}=N^{1+a-p(1s)}v(a)^{1-m(1s)}E^{1s}+\dots
\end{equation} 
gives the condition $1+a-p(1s)\leq 0$. Note that for a fixed $a$ this condition permits  arbitrarily large $p(1s)$, at the price of removing terms from the large $N$ limit of the algebra. A general lesson is that the critical values when an inequality becomes an equality produces a more general limit algebra than when the inequality is strictly satisfied. Thus, we are primarily interested in cases on the boundary of the allowed region, or even its edges or corners. Relaxing a boundary equality to an inequality only gives a special case of a more general large $N$ limit. A few of the constraints we obtain are
\begin{equation}
\begin{split}
&1+a \leq p(1s)\\
& 1 \leq p(1t) \\
&2+a \leq p(2s) \\
&2 \leq p(2t) \\
&a \leq p(2t)\\
& 1+p(1s) \leq 2p(1t).
\end{split}
\end{equation}
The constraints give the condition  $0\leq a\leq 2$ for a solution to exist for a well-defined algebra. 
The most general solutions to the constraints are 
\begin{equation}\label{eq_p}
\begin{split}
&p(1s)=1+a \\
&p(1t)=1+\frac{a}{2} \\
&p(2s) = 2+a \\
& p(2t)=2,
\end{split}
\end{equation}
which hold for $0\leq a\leq 2$. Other solutions can be seen as a special cases of these.

Equation (\ref{eq_p}) gives the coefficients $n(k)$ and $m(k)$, which have no $a$-dependence, and through (\ref{eq_MN2}) and (\ref{eq_lgLG}) we obtain the scalings for the couplings
\begin{equation} \label{eq_scalingL}
\begin{split}
\lambda_{1s} &= \frac{\Lambda_{1s}}{MN} = \frac{\lambda_{1S}}{MN} \\
\lambda_{1t} &= \frac{\Lambda_{1t}}{\sqrt{M}N} = \frac{\lambda_{1T}}{\sqrt{M}N} \\
\lambda_{2s} &= \frac{\Lambda_{2s}}{MN^2} =  \frac{\lambda_{2S}}{MN^2} \\
\lambda_{2t} &= \frac{\Lambda_{2t}}{N^2}= \frac{\lambda_{2T}}{N^2}. \\
\end{split}
\end{equation}
These are the scalings of the couplings we will use for the remainder of the analysis of our model. 
We have introduced rescaled basis elements and couplings by conveniently denoting them 
with capital superscripts and subscripts, ie $E^k \rightarrow e^K$ and $\Lambda_k \rightarrow \lambda_K$ where $K=1S,1T,2S,2T$. In fact, these rescaled couplings are also the couplings one obtains in multi-matrix models by applying the combinatorial prescriptions in \cite{Azeyanagi:2017mre}.

We can now take three different large $N$ limits, depending on the $N$ and $M$ relation (\ref{eq_MN}), which are
\begin{itemize}
    \item \textbf{Case $a=2$:} a `multi-matrix' limit where $M=v(2)N^2$ and $v(2)\equiv v$ is a constant free parameter,
    \item \textbf{Case $0<a<2$:} an intermediate case where $M=v(a) N^a$ and $v(a)$ is a constant free parameter,
    \item \textbf{Case $a=0$:} the regular 't Hooft limit where $M=v(0)$ is a constant free parameter.
\end{itemize}

\subsection{One-loop algebras at large $N$ and $M$}\label{sec_LargeN}
We calculate the algebra for each large $N$ case by first calculating the multiplication matrix and then taking the large $N$ limit and discarding subleading terms. 
In this section we first discuss the algebras for the three cases respectively, and then the subalgebras and ideals for all three cases (shown in table \ref{T1}).

\paragraph{Case $a=2$}\label{sec_a2}
First we consider the algebra for the case $a=2$, where $M=vN^2$ and $v$ is a free parameter, with the multiplication table shown in table \ref{Ta2}. 
We note that the free parameter $v$ appears in the products and that for $v=0$ the algebra reduces to the algebra for the case $0<a<2$.
\begin{table}[tbp] 
\begin{center}
\begin{tabular}{ |c|c|c|c|c| } 
 \hline
 $\diamond$   & $e^{1S}$  & $e^{1T}$  & $e^{2S}$  & $e^{2T}$\\
\hline
  $e^{1S}$  & $\frac{1}{2} e^{1 S} + \frac{1}{2}e^{2 S}$  &  $0$ & $e^{2 S}$  & $2v e^{1 S}+2e^{2 S}$\\
 \hline
$e^{1T}$   & $0$  & $\frac{1-v}{2} e^{1 S} + \frac{v}{4}e^{2S}+ \frac{1}{8} e^{2 T} $   & $0$  & $2v e^{1 T}$\\
\hline
$e^{2S}$   &  $e^{2 S}$ & $0$ & $e^{2 S}$  & $(2+2v) e^{2 S}$\\
 \hline
$e^{2T}$   &  $2v e^{1 S}+2e^{2 S}$ & $2v e^{1 T}$  &  $(2+2v) e^{2 S}$ & $12ve^{2S} + (2+2v)e^{2 T}$ \\
 \hline
\end{tabular} 
\end{center}
\caption{\label{Ta2} Large $N$ algebra for the case $a=2$ where $M=vN^2$ and $v$ is a constant.}
\end{table}

An example of what we can observe from the algebra is that the single-trace couplings induce running in the double-trace couplings. This is due to the ST-ST products yielding both ST and DT terms, while DT-DT products only yield DT terms.

\paragraph{Case $0<a<2$} \label{sec_0a2}
The algebra for the case $0<a<2$, where $M=v(a)N^a$ and $v(a)$ is a constant, is shown in table \ref{T0a2}. 
In this case the algebra does not have a free parameter since $v(a)$ does not appear in any of the products. It is also a common special case of the algebras for the cases $a=0$ and $a=2$, since taking $M\rightarrow \infty$ for the case $a=0$, and taking $v\rightarrow 0$ for the case $a=2$ both yield this algebra. 
\begin{table}[ht] 
\begin{center}
\begin{tabular}{ |c|c|c|c|c| } 
 \hline
 $\diamond$   & $e^{1S}$  & $e^{1T}$  & $e^{2S}$  & $e^{2T}$\\
\hline 
  $e^{1S}$  & $\frac{1}{2} e^{1 S} + \frac{1}{2}e^{2 S}$  & $0$ & $e^{2 S}$  & $2e^{2 S}$\\
 \hline
$e^{1T}$   & $0$  & $\frac{1}{2} e^{1 S} + \frac{1}{8} e^{2 T}$   & $0$  & $0$\\
\hline
$e^{2S}$   &  $e^{2 S}$ & $0$ & $e^{2 S}$  & $2e^{2 S}$\\
 \hline
$e^{2T}$   &  $2e^{2 S}$ & $0$  &  $2e^{2 S}$ & $2e^{2 T}$ \\
 \hline
\end{tabular} 
\end{center}
\caption{\label{T0a2} Large $N$ algebra for the case $0<a<2$.} 
\end{table}

We note that the single-trace element $e^{1T}$ is not generated in any of the products, which means that all structure constants $C^{mn}_{1T}=0$. This implies that $\beta_{1T}=0$ (to one-loop order).

\paragraph{Case $a=0$} \label{sec_a0}
Finally, we consider the algebra for the regular 't Hooft large $N$ limit, where $M$ is a constant free parameter, shown in table \ref{Ta0}. This algebra corresponds to a rescaled (with respect to $M$) version of the large $N$ algebra considered in \cite{Flodgren:2023lyl}.\footnote{In \cite{Flodgren:2023lyl} another large $N$ scaling of the basis elements/couplings was used when calculating the multiplication matrix. However, the algebras in both cases have the same invariant properties.} 
\begin{table}[ht] 
\begin{center}
\begin{tabular}{ |c|m{3.5cm}|m{3.5cm}|m{2cm}|m{2cm}| } 
 \hline
  $\diamond$  & \hfil$e^{1S}$  & \hfil$e^{1T}$  & \hfil$e^{2S}$  & \hfil$e^{2T}$\\
\hline 
  $e^{1S}$  & \hfil$(\frac{1}{2}+\frac{3}{2M}) e^{1 S} + \newline (\frac{1}{2}+\frac{3}{2M}) e^{2 S} +\frac{1}{2M^2} e^{2 T}$  
  &  \hfil$\frac{1}{2\sqrt{M}} e^{1 S}+\frac{1}{M}e^{1 T} \newline+ \frac{1}{2\sqrt{M}} e^{2 S}+\frac{1}{2M^{3/2}} e^{2 T}$ 
  & \hfil$(1+\frac{1}{M}) e^{2 S}$  & \hfil$2 e^{2 S}+\frac{1}{M}e^{2 T}$\\
 \hline
$e^{1T}$   & \hfil$\frac{1}{2\sqrt{M}} e^{1 S}+\frac{1}{M}e^{1 T} \newline+ \frac{1}{2\sqrt{M}} e^{2 S}+\frac{1}{2M^{3/2}} e^{2 T}$  
& \hfil$\frac{1}{2} e^{1 S} + (\frac{1}{8}+\frac{1}{4M}) e^{2 T}$  
& \hfil$\frac{1}{\sqrt{M}}e^{2 S}$  
& \hfil$\frac{1}{\sqrt{M}}e^{2 T}$\\
\hline
$e^{2S}$   &  \hfil$(1+\frac{1}{M}) e^{2 S}$ 
& \hfil$\frac{1}{\sqrt{M}}e^{2 S}$ 
& \hfil$e^{2 S}$  
& \hfil$2e^{2 S}$\\
 \hline
$e^{2T}$   &  \hfil$2 e^{2 S}+\frac{1}{M}e^{2 T}$ 
& \hfil$\frac{1}{\sqrt{M}}e^{2 T}$  
&  \hfil$2e^{2 S}$ 
& \hfil$2e^{2 T}$ \\
 \hline
\end{tabular} 
\end{center}
\caption{\label{Ta0} Large $N$ algebra for the case $a=0$ where $M$ is a constant.}
\end{table}

We note that the parameter $M$ appears in the products and that in the limit $M \rightarrow \infty$ the algebra reduces to the algebra of the $0<a<2$ case.

\paragraph{Subalgebras and quotient algebras}\label{sec_T1}
The algebra gives us information about the RG equations and flows. Now we focus on what the subalgebras of an algebra can convey. Notation-wise, we use the brackets $\{e^i,\dots, e^k
\}$ to signify a linear space spanned by the elements in the brackets. For example, the full algebra for any of the cases is denoted $A=\{e^{1S},e^{1T},e^{2S},e^{2T}\}$. We will here consider linear spans of one or several of these symmetry-respecting basis elements, and find subspaces which are closed under multiplication. 

We have seen that an algebra of couplings is associated with renormalisable scalar theories at one-loop order. A subalgebra of an algebra is a linear subspace of the algebra which is closed under multiplication and thus an algebra. The scalar subtheory with couplings in the subalgebra is renormalisable at one-loop order just as the full theory, due to the closure of the multiplication. 

An ideal is a more restrictive concept than a subalgebra. An ideal is a subalgebra with the extra requirement that the product of \emph{any} element of the algebra with an element in the ideal belongs to the ideal. This allows for the definition of a quotient algebra - obtained by modding out the algebra by an ideal. This can be done consistently since multiplying any element by elements of the ideal keeps the product in the ideal, thus not affecting the quotient algebra.

Writing general elements in the algebra as $\lambda_m q^m + I$ where $I$ stands for any element in the ideal $I$ and $q^m$ is a representative of the quotient algebra, we find that $(\kappa_m q^m + I) \diamond (\lambda_n q^n + I) = \kappa_m \lambda_n C^{mn}_k q^k + I$. We can simply disregard the components of couplings in the ideal $I$. For the symmetry-respecting bases and the symmetry-respecting ideals we consider in this section, the $q^k$ can simply be chosen to be basis elements not contained in the ideal $I$. Then $\beta_q = C_q^{pr} \lambda_p \lambda_r$ where the indices $p,q,r$ run over the labels which do not belong to the ideal. Thus, the RG equations for the coupling constants in the quotient algebra form a closed dynamical system, independent of the coupling constants in the ideal.

In short, the subalgebras describe which elements induce other elements, while quotient algebras tell us about the structure of the one-loop RG equations. 
Table \ref{T1} shows some of the subalgebras, ideals and their corresponding quotient algebras for the three types of limit algebras. There are possibly more subalgebras and ideals which cannot be described simply as linear spans of our basis elements.
\begin{table} 
\begin{center}
\begin{tabular}{ |m{2.5cm}|m{2.75cm}|m{2.75cm}|m{2.75cm}| } 
 \hline
 & \hfil \textbf{a=0} & \hfil \textbf{0$<$a$<$2} & \hfil \textbf{a=2}\\ 
 \hline
 \textbf{Subalgebras} & $\{e^{2S}\}$, $\{e^{2T}\}$, \newline $\{e^{2S},e^{2T}\}$, $\{e^{1S},e^{2S}$, $e^{2T}\}$ 
 & $\{e^{2S}\}$, $\{e^{2T}\}$, \newline $\{e^{1S}, e^{2S}\}$, $\{e^{2S},e^{2T}\}$, $\{e^{1S},e^{2S},e^{2T}\}$ 
 & $\{e^{2S}\}$, \newline $\{e^{1S}, e^{2S}\}$, $\{e^{2S},e^{2T}\}$, $\{e^{1S},e^{2S},e^{2T}\}$\\ 
 \hline
 \textbf{Ideals} & $\{e^{2S}\}$, $\{e^{2S},e^{2T}\}$ 
 & $\{e^{2S}\}$, $\{e^{1S}, e^{2S}\}$, $\{e^{2S},e^{2T}\}$, $\{e^{1S},e^{2S},e^{2T}\}$ 
 & $\{e^{2S}\}$, $\{e^{1S}, e^{2S}\}$\\ 
 \hline
   \textbf{Quotient \newline algebras} & $A/\{e^{2S}\}$, $A/\{e^{2S},e^{2T}\}$ 
   & $A/\{e^{2S}\}$, $A/\{e^{1S},e^{2S}\}$,  $A/\{e^{2S},e^{2T}\}$, $A/\{e^{1S},e^{2S},e^{2T}\}$ 
   & $A/\{e^{2S}\}$, $A/\{e^{1S},e^{2S}\}$\\ 
 \hline
\end{tabular} 
\end{center}
\caption{\label{T1} Table of the subalgebras, ideals and quotient algebras of the ideals for the three algebra cases. This table only contains symmetry-respecting subspaces (ie subspaces spanned by one or several basis elements of the symmetry-respecting basis).}
\end{table}

Let us look at an example. In the $a=0$ case we have a double-trace ideal $I^2=\{e^{2S},e^{2T}\}$. The quotient algebra $A/\{e^{2S},e^{2T}\}$ is then given by the multiplication table of the single-trace elements $e^{1S}$ and $e^{1T}$ with the double-trace terms modded out. This tells us that the beta functions $\beta_{1S}$ and $\beta_{1T}$ 
are independent of the double-trace couplings $\lambda_{2S}$ and $\lambda_{2T}$. Therefore, in order to find the RG flow for the full system we can start by solving the single-trace RG equations. 
By considering the rest of the ideals of this algebra, ie $\{e^{2S}\}$, we find that the natural order in which to solve the full system of RG equations is to first solve $ \dot{\lambda}_{1S}=\beta_{1S}$ and $\dot{\lambda}_{1T}=\beta_{1T}$ as a system of equations, then $\dot{\lambda}_{2T}=\beta_{2T}$, and finally $\dot{\lambda}_{2S}=\beta_{2S}$.

The algebra for the case $0<a<2$ has the most ideals of the three cases. This implies more independent RG equations for the couplings. Here the ideals imply that the order of solving the RG equations is first $\dot{\lambda}_{1T}=\beta_{1T}=0$, second $\dot{\lambda}_{1S}=\beta_{1S}$ and $\dot{\lambda}_{2T}=\beta_{2T}$ as two independent equations, and third $\dot{\lambda}_{2S}=\beta_{2S}$. 

The final case $a=2$ has a similar hierarchy to the $a=0$ case, since the only difference in the algebra is that instead of the double-trace ideal $I^2$ it has a scalar-factor ideal $I^S=\{e^{1S},e^{2S}\}$. The order is first $\dot{\lambda}_{1T}=\beta_{1T}$ and $\dot{\lambda}_{2T}=\beta_{2T}$ as a system of equations, second $\dot{\lambda}_{1S}=\beta_{1S}$ and third $\dot{\lambda}_{2S}=\beta_{2S}$. 

The algebra before we take the large $N$ limit, given by the equations (\ref{eq_Nprod1}-\ref{eq_Nprod3}), has the subalgebras $\{e^{2s}\}$ and $\{e^{2s},e^{2t}\}$, but no ideals. These subalgebras correspond to the rescaled large $N$ subalgebras $\{e^{2S}\}$ and $\{e^{2S},e^{2T}\}$, which exist for all three cases in table \ref{T1}. The lack of ideals tells us that all the beta functions \textit{can} depend on all the couplings, but there is no guarantee that they do.

\section{Properties of the algebras}\label{sec_3}

The idea that the RG flow (at least for weak coupling) is captured by an algebra prompts us to characterise this algebra as thoroughly as possible. We show that special elements, idempotents and nilpotents, characterise the algebras. Furthermore, for each idempotent, its linear action on the algebra defines an eigenvalue problem and a decomposition into eigenspaces called a Peirce decomposition. These ideas are then applied to the large $N$ algebras of section \ref{sec_LargeN}. Finally, we discuss the physical interpretation of all these concepts in terms of the RG flow.

\subsection{Special elements of the algebra and Peirce decompositions}
An algebra can be characterised by its structure constants, which are given in the multiplication table. This description, however, depends on the choice of basis. There are usually a number of special elements, which are defined invariantly by specific properties. Similarly, the associated systems of non-linear ordinary differential equations, like the RG equations of this paper, have a number of especially simple solutions. They are directly related to the special algebra elements. In the spirit of an algebraic description, we first recall algebra elements purely defined by their products with themselves. Idempotent elements, or idempotents $\textbf{c}$ 
of an algebra fulfil
\begin{equation}
\textbf{c}^2\equiv \textbf{c}\diamond \textbf{c} = \textbf{c},
\end{equation}
while nilpotent elements, or nilpotents $\textbf{n}$, fulfil
\begin{equation}
\textbf{n}^2\equiv \textbf{n}\diamond \textbf{n} = 0.
\end{equation}

\paragraph{Peirce decompositions}\label{sec_PD}
The idempotents of a non-associative algebra characterise the algebra, and an even finer characterisation is obtained through their so called Peirce decompositions. As emphasised in \cite{Krasnov_2018} many invariant properties can be recovered from the Peirce spectrum of the algebra. 

The Peirce spectrum $\sigma(\textbf{x})$ is the eigenvalue spectrum of the linear multiplication operator $L_{\textbf{x}}$, where $\textbf{x}$ is a linear combination of basis elements for an algebra. The linear multiplication operators
\begin{equation}
L_{\textbf{x}}: L_{\textbf{x}}\textbf{y} \equiv \textbf{x}\diamond \textbf{y} = \textbf{y}\diamond \textbf{x}
\end{equation}
form an associative algebra acting on the vector space $A$.  
The Peirce spectra of non-zero idempotent elements $\textbf{c}: \textbf{c}^2=\textbf{c}$ have eigenvalues called Peirce numbers. The finite set of Peirce numbers are a useful way of characterizing the algebra. 
A Peirce decomposition is the decomposition of the algebra into the eigenspaces of the $L_{\textbf{c}}$ operators for an idempotent.

Concretely, the idempotents, the Peirce decomposition and the Peirce spectra add fine structure to the structure of ideals and quotient algebras we have already uncovered. From table  \ref{T1} the smallest closed dynamical subsystems, given by quotient algebras, are two-dimensional for $a = 2$ and $a = 0$ and one-dimensional for $0 < a < 2$. The Peirce decomposition does not give any new information in one-dimensional systems, but we will see its usefulness in the cases $a=2$ and $a=0$ below.

\paragraph{Peirce decomposition for $a=2$}\label{sec_Pa2}
Let us consider the Peirce decomposition for the algebra of case $a=2$, ie the large $N$ limit where $M=vN^2$ with $v$ constant. This case has the 2D ideal $I^{S}=\{e^{1S},e^{2S}\}$, and we will calculate the Peirce decomposition of its quotient algebra $A/I^S$ which defines a closed dynamical system. 

The quotient algebra $A/I^S$ has a multiplication table given by table \ref{Tp1}. 
To calculate the idempotents, we look for linear combinations of the basis elements $\textbf{c}=c_Ie^I$, where $I=1T,2T$, that are idempotent $\textbf{c}=\textbf{c}\diamond \textbf{c}.$ There are three idempotents
\begin{equation}\label{eq_a2idempotents}
\begin{aligned}
\textbf{c}^0(v) &= \frac{1}{2+2v}e^{2T} \\
\textbf{c}^{\pm}(v) &= \pm \frac{\sqrt{v-1}}{v}e^{1T}+\frac{1}{4v}e^{2T}.
\end{aligned}
\end{equation}
The two idempotents $\textbf{c}^{\pm}(v)$ are real for $v\geq 1$, while $\textbf{c}^{0}(v)$ is real for all $v$. 
The complex idempotents however diverge in the limit $v\to 0$. If the divergence is scaled out one finds nilpotent limiting elements $\textbf{n}^{\pm}(v)=v \textbf{c}^{\pm}(v)$ as $v\to 0$. Simultaneously, the space $\{\textbf{c}^{0}(v)\}$ becomes a 1D ideal (for the 2D quotient algebra, not the full algebra) as can be seen from table \ref{Tp1}. 
After modding out $\{\textbf{c}^{0}(v)\}$ only the imaginary parts of two nilpotents $\textbf{n}^{\pm}(v)$ survive. They are both proportional to $\{e^{1T}\}$ and thus only span a 1D ideal. Again referring to table \ref{Tp1}, we see that the nilpotency is preserved in the quotient.
\begin{table} 
\begin{center}
\begin{tabular}{ |c|c|c| } 
 \hline
  $\diamond$  & $e^{1T}$  & $e^{2T}$\\
\hline 
  $e^{1T}$  & $\frac{1}{8}e^{2T}$  & $2ve^{1T}$\\
 \hline
$e^{2T}$    & $2ve^{1 T}$  & $(2+2v)e^{2T}$\\
 \hline
\end{tabular} 
\end{center}
\caption{\label{Tp1} Table of the $a=2$ case quotient algebra $A/I^{S}$.}
\end{table}

In order to show the usefulness of a Peirce decomposition we are also interested in the Peirce numbers which are invariants of the algebra. They are eigenvalues associated to the idempotents. We can see why they are invariants as follows. While explicit expressions for idempotents depend on the basis, the idempotents themselves are invariants because their defining equations are basis independent. Thus, the Peirce numbers are also invariants. To calculate the Peirce numbers we consider the linear multiplication operators for the idempotents $L_{\textbf{c}}$. The operator $L_{\textbf{c}}$ is a $2\times 2$ matrix for $A/I^S$. One eigenvalue of each $L_{\textbf{c}}$ is always equal to one, since $L_{\textbf{c}}\textbf{c}=\textbf{c}\diamond \textbf{c}=\textbf{c}=1\textbf{c}$. We call these eigenvalues trivial. The operators $L_{\textbf{c}}$ for the idempotents (\ref{eq_a2idempotents}) are
\begin{equation}
\begin{aligned}
L_{c^0} &= 
\left(
\begin{array}{cc}
\frac{v}{1+v}  & 0  \\
0  &  1
\end{array}
\right)
\end{aligned}
\end{equation}
and 
\begin{equation}
\begin{aligned}
L_{c^{\pm}} &= 
\left(
\begin{array}{cc}
\frac{1}{2}  & \pm2\sqrt{v-1}  \\
 \pm\frac{\sqrt{v-1}}{8v}  & \frac{(1+v)}{2v}
\end{array}
\right). \\
\end{aligned}
\end{equation}
The respective non-trivial Peirce numbers are 
\begin{equation} \label{eq_PNa2}
\begin{aligned}
\mu_0(v)=\frac{v}{1+v}, \mu_+(v)=\mu_-(v)=\frac{1}{2v}. 
\end{aligned}
\end{equation}
The Peirce numbers $0$ and $1/2$ indicate special properties of the algebra, eg a merger  
of several idempotents. Here we note that for $v=1$ all Peirce numbers are $1/2$ and all idempotents are equal.

\paragraph{Peirce decomposition for $a=0$}\label{sec_Pa0}
The calculations for the large $N$ case where $a=0$ and $M$ is a finite free parameter are similar to the ones for the $a=2$ case. 

This case has a 2D ideal $I^2=\{e^{2S},e^{2T}\}$ with quotient algebra $A/I^2$, which is given in table \ref{Tp2}. 
The idempotents for this quotient algebra are 
\begin{equation}\label{eq_idempotentsa0}
\begin{aligned}
\textbf{c}^0(M) &= \frac{2M}{M+3}e^{1S} \\
\textbf{c}^{\pm}(M) &= \frac{M}{2}e^{1S} +(- \frac{1}{2} \pm \frac{\sqrt{2-M}}{2})\sqrt{M}e^{1T}.
\end{aligned}
\end{equation}
$\textbf{c}^{\pm}(M)$ are real for $M \leq 2$ and $\textbf{c}^0(M)$ is real for all $M$.  In the limit $M\to \infty$ the complex idempotents again diverge, and become nilpotents if they are rescaled. In fact, this limit yields the same $0<a<2$ case as the previous $v\to 0$ limit for $a=2$, again producing a nilpotent 1D ideal corresponding to the limit of the imaginary part of the complex idempotents $\textbf{c}^{\pm}(M)$.
\begin{table} 
\begin{center}
\begin{tabular}{ |c|c|c| } 
 \hline
  $\diamond$  &  $e^{1S}$  & $e^{1T}$\\
\hline 
  $e^{1S}$  & $\frac{M+3}{2M}e^{1S}$  & $\frac{1}{2\sqrt{M}}e^{1S}+\frac{1}{M}e^{1T}$\\
 \hline
$e^{1T}$    & $\frac{1}{2\sqrt{M}}e^{1S}+\frac{1}{M}e^{1T}$  & $\frac{1}{2}e^{1S}$\\
 \hline
\end{tabular} 
\end{center}
\caption{\label{Tp2} Table of the $a=0$ case quotient algebra $A/I^{2}$.}
\end{table}

The linear multiplication operators in the 2D simple ideal for $a=0$ are
\begin{equation}
\begin{aligned}
L_{c^0} &= 
\left(
\begin{array}{cc}
1 & \frac{\sqrt{M}}{M+3}  \\
0  &  \frac{2}{M+3}
\end{array}
\right)
\end{aligned}
\end{equation}
and 
\begin{equation}
\begin{aligned}
 L_{c^{\pm}} &= 
\left(
\begin{array}{cc}
\frac{M+2}{4}\pm \frac{\sqrt{2-M}}{4}  & \pm\frac{\sqrt{(2-M)M}}{4}  \\
(-\frac{1}{2}\pm\frac{\sqrt{2-M}}{2})\frac{1}{\sqrt{M}} & 1/2
\end{array}
\right), \\
\end{aligned}
\end{equation}
with the respective non-trivial Peirce numbers 
\begin{equation}\label{eq_PNa0}
\begin{aligned}
\mu_0(M)=\frac{2}{M+3}, \mu_{\pm}(M)=\frac{1}{4}(M \pm \sqrt{2-M}).
\end{aligned}
\end{equation}
$\mu_0(M)$ is real for all $M$ and $\mu_{\pm}(M)$ are real for $M\leq2$, like the idempotents. 

Here the Peirce number $1/2$ appears for two values of $M$. 
For $M=1$ the idempotents $\textbf{c}^0(1)=\textbf{c}^+(1)$ and their Peirce numbers are $1/2$, and similarly for $M=2$ where $\textbf{c}^+(1)=\textbf{c}^-(1)$ with Peirce numbers $1/2$.

\subsection{Physical interpretation}

An idempotent $\textbf{c}$ is a special coupling which reproduces itself when squared. The corresponding RG flow will just decrease or increase this coupling, and will not induce other couplings.
Therefore, the idempotents correspond to one-loop RG flows in one-dimensional linear subspaces of couplings, called \emph{ray solutions}. The Peirce decompositions in their turn characterise the small perturbations around the ray solutions. A Peirce number $\mu=0$ simply means that the perturbation away from the ray solution defined by the idempotent neither grows nor diminishes, as is expected for directions orthogonal to an ideal containing the idempotent. The trivial Peirce number $\mu=1$ corresponds to an exponential evolution of the perturbation with the same exponent as the ray solution of the idempotent. Finally, the Peirce number $\mu=\frac{1}{2}$ is also special: it corresponds to the exponential with precisely half of the exponent of the ray solution. The Peirce numbers actually appeared in the Riccati equations for the spinning top analysed by Sonya Kowalevski in the 1890s \cite{10.1007/BF02413316}. In the ODE context they are thus Kowalevski exponents \cite{math11081790}. 

We find that the RG flow is characterised by ray solutions and their Kowalevski exponents, ie the idempotents and their Peirce spectra. This picture was recently developed by Krasnov \cite{math11081790}, especially in 2d flows, where the rays separate the plane into sectors. Then the topology of the flow is given by the sequence of sectors and the Kowalevski exponents of the bounding ray solutions for each sector. 

Krasnov classifies four types of rays along the idempotents based on their Peirce numbers \cite{math11081790}. Rays are lines along the directions of the idempotents while the idempotents themselves are vectors in the basis space. 
Rays are called exceptional if their Peirce number spectra contains $0$ or $1/2$. As we have mentioned, exceptional type rays seem to indicate a merger of several idempotents. 
In two-dimensional algebras a sector is the domain bounded by two subsequent rays and there are six non-exceptional types of sectors. 
We will discuss rays/idempotents and their corresponding Peirce numbers for the RG flows but we will not use the full power of this classification in this paper.

\section{One-loop RG flow at large $N$}\label{sec_4}
The following analysis of the beta functions is based on the quotient algebras, idempotents and Peirce numbers calculated in sections \ref{sec_LargeN} and \ref{sec_PD}. We comment on what these objects imply for the RG flow for the cases $a=2$, $0<a<2$ and $a=0$.

\subsection{Case $a=2$}
Let us consider the full system of one-loop beta functions for the case $a=2$, where $M=vN^2$ and $v$ is a constant. The beta functions are
\begin{equation} \label{eq_beta2}
\begin{aligned}
\beta_{1 S}= & \frac{1}{32\pi^2}(\lambda_{1S}^2+(1-v)\lambda_{1T}^2+8v\lambda_{1S}\lambda_{2T}) \\
\beta_{1 T}= & \frac{1}{4\pi^2}v\lambda_{1T}\lambda_{2T} \\
\beta_{2 S}= & \frac{1}{64\pi^2}(2\lambda_{1S}^2+8\lambda_{1S}(\lambda_{2S}+2\lambda_{2T}) +4\lambda_{2S}(\lambda_{2S}+4\lambda_{2T}) +v(\lambda_{1T}^2+16\lambda_{2T}(\lambda_{2S}+3\lambda_{2T})) ) \\
\beta_{2 T}= & \frac{1}{128\pi^2}(\lambda_{1T}^2+16(1+v)\lambda_{2T}^2).
\end{aligned}
\end{equation}
The only fixed point for this system of equations is the origin.

In table \ref{T1} we show four subalgebras for this case. Each subalgebra forms a renormalisable theory on its own. For example, the subalgebra $\{e^{2S},e^{2T}\}$ implies that setting $\lambda_{1S}=\lambda_{1T}=0$ leads to the only  non-zero beta functions being $\beta_{2S}$ and $\beta_{2T}$. Equation (\ref{eq_beta2}) shows that this is the case.  

In section \ref{sec_LargeN} we show that the ideals, via their quotient algebras, imply that the hierarchy for solving the RG equations is first $\dot{\lambda}_{1T}=\beta_{1T}$ and $\dot{\lambda}_{2T}=\beta_{2T}$ as a system of equations, second $\dot{\lambda}_{1S}=\beta_{1S}$ and third $\dot{\lambda}_{2S}=\beta_{2S}$. Equation (\ref{eq_beta2}) shows this hierarchy via which beta functions depend of which couplings. For example, $\beta_{2S}$ must be solved last since it depends on all four couplings.

\begin{figure}
     \centering
     \begin{subfigure}[b]{0.45\textwidth}
         \centering
         \includegraphics[width=\textwidth]{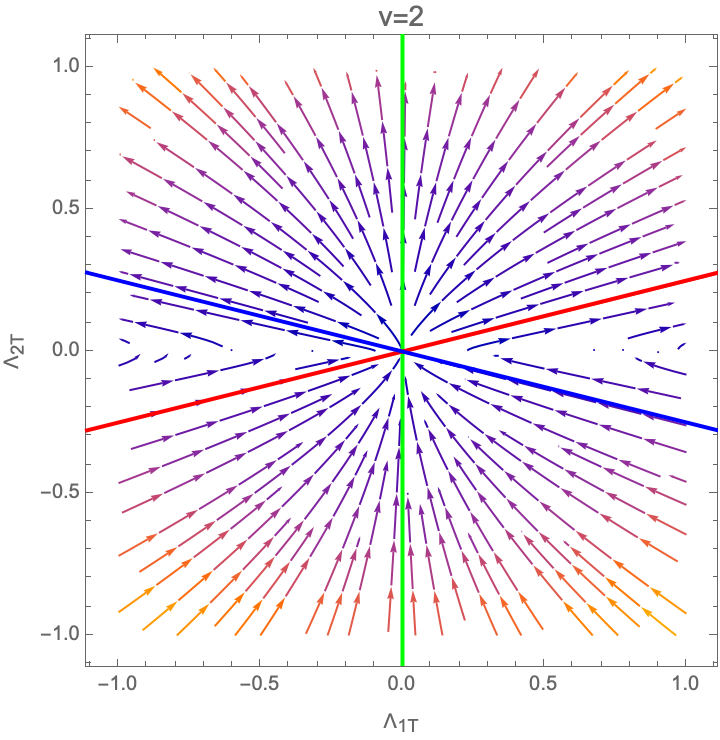}
         \caption{$v=2$}
         \label{fig:smallv}
     \end{subfigure}
     \hfill
     \begin{subfigure}[b]{0.45\textwidth}
         \centering
         \includegraphics[width=\textwidth]{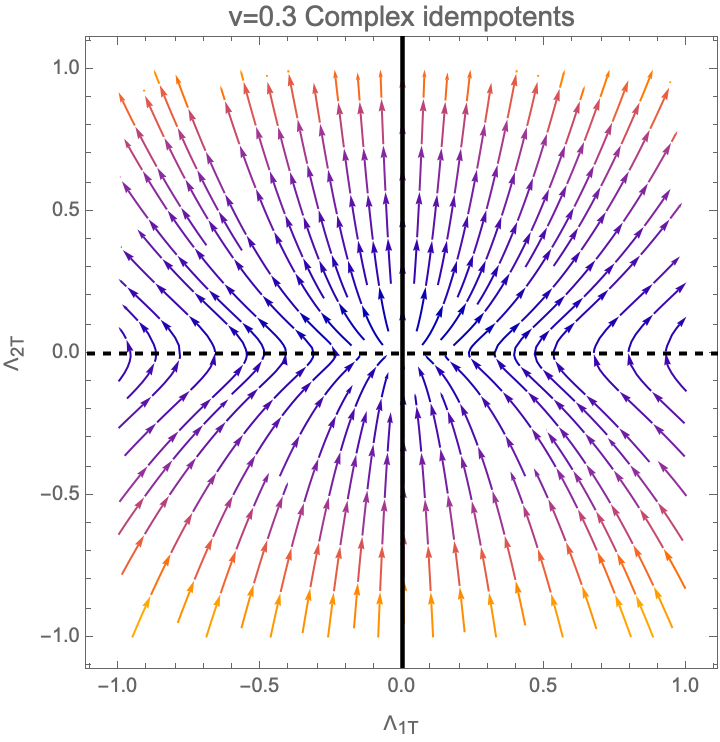}
         \caption{$v=0.3$}
         \label{fig:smallerv}
     \end{subfigure}
     \hfill
        \caption{
       RG flows for the case $a=2$. The figures show the 2D flows in the closed dynamical system that is the $(\lambda_{1T},\lambda_{2T})$-space. The color of the flow lines represents the flow velocity, where purple is a lower velocity than yellow. \textbf{(a)} For $v=2$ the ray solutions of the idempotents $\textbf{c}^0(v)$, $\textbf{c}^+(v)$ and $\textbf{c}^-(v)$ are green, red and blue lines respectively. 
        As $v$ decreases the red and blue rays move away from the $\lambda_{1T}$-axis towards the $\lambda_{2T}$-axis. 
        \textbf{(b)} For  $v=0.3$ only
        $\textbf{c}^0(v)$ is real. $\textbf{c}^0(v)$, $\Re(\textbf{c}^+(v))$ and $\Re(\textbf{c}^-(v)$) overlap along the $\lambda_{1T}$-axis and are shown as a black line. $\Im(\textbf{c}^+(v))$ and $\Im(\textbf{c}^-(v))$ overlap along the $\lambda_{2T}$-axis and are represented by the dashed black line. As $v$ changes in the range $0<v<1$ these lines do not change, but $\textbf{c}^+(v)$ and $\textbf{c}^-(v)$ change magnitude and phase.}
        \label{fig:1T2T}
\end{figure}

Next, we consider the RG flow in the closed dynamical system that is the coupling space of the quotient algebra of the 2D ideal $A/I^S$. It corresponds to the couplings space $(\Lambda_{1T},\Lambda_{2T})$. 
This quotient algebra has three idempotents given by (\ref{eq_a2idempotents}), the directions of which we should see in the RG flow, and the corresponding Peirce numbers (\ref{eq_PNa2}). 
The idempotents are all real for $v\geq 1$. 
At $v=1$ the idempotents are all equal, ie the rays merge, $\textbf{c}^{+}=\textbf{c}^{-}=\textbf{c}^{0}$ and their Peirce numbers are $1/2$, which indicates exceptional type rays. 
For $v<1$ two of the idempotents are complex, however, their Peirce numbers are real for all $v$. 
The RG flow in the $(\lambda_{1T},\lambda_{2T})$-space is closed and can be seen in figure \ref{fig:1T2T} for $v=2$ and $v=0.3$, along with the idempotents. There is only one actual real ray solution for $v<1$, and some separatrices, real RG trajectories separating the solutions which reach the origin from those that do not, switch from being rays to being curved when two idempotents collide and get imaginary parts for $v<1$.
\subsection{Case $0<a<2$}

The one-loop beta functions for the large $N$ limit where $M=v(a)N^a$, with $0<a<2$ and $v(a)$ constant, are 
\begin{equation} \label{eq_beta0a2}
\begin{aligned}
\beta_{1 S}= & \frac{1}{32\pi^2}(\lambda_{1S}^2+\lambda_{1T}^2) \\
\beta_{1 T}= & 0 \\
\beta_{2 S}= & \frac{1}{32\pi^2}(\lambda_{1S}^2+4\lambda_{1S}\lambda_{2S}+2\lambda_{2S}^2+8(\lambda_{1S}+\lambda_{2S})\lambda_{2T}) \\
\beta_{2 T}= & \frac{1}{128\pi^2}(\lambda_{1T}^2+16\lambda_{2T}^2).
\end{aligned}
\end{equation}
The parameter $v(a)$ does not appear in the beta functions, meaning that there is no free parameter in this limit and the RG flow does not depend on the value of $M$. In contrast to the other cases, which only have the trivial fixed point, here we have four variables but only three fixed point equations, implying we have a line of fixed points. This line, however, is complex. 
Choosing $\lambda_{1T}$ to be purely imaginary, the one-loop fixed point equations determine the other couplings to be completely real and functions of $\lambda_{1T}$. As discussed in section \ref{conclusion}, this line of fixed points does not survive two-loop corrections. 

This case has four symmetry-respecting ideals, see table \ref{T1}, and respective quotient algebras. The quotient algebras imply that the order for solving the RG equations is first $\dot{\lambda}_{1T}=\beta_{1T}$, second $\dot{\lambda}_{1S}=\beta_{1S}$ and $\dot{\lambda}_{2T}=\beta_{2T}$ as two independent equations, and third $\dot{\lambda}_{2S}=\beta_{2S}$. 

\subsection{Case $a=0$}

The beta functions for the large $N$ limit with finite $M$ are 
\begin{equation} \label{eq_beta0}
\begin{aligned}
\beta_{1 S}= & \frac{1}{32\pi^2M}((3+M)\lambda_{1S}^2 + 2\sqrt{M}\lambda_{1S}\lambda_{1T} + M \lambda_{1T}^2 ) \\
\beta_{1 T}= & \frac{1}{8\pi^2M}\lambda_{1S}\lambda_{1T} \\
\beta_{2 S}= & \frac{1}{32\pi^2M}((3+M)\lambda_{1S}^2 +2\sqrt{M}\lambda_{1S}\lambda_{1T} +2\lambda_{2S}(2(1+M)\lambda_{1S} +2\sqrt{M}\lambda_{1T}+M\lambda_{2S}) \\
&+ 8M(\lambda_{1S}+\lambda_{2S})\lambda_{2T} ) \\
\beta_{2 T}= & \frac{1}{128\pi^2M^2}(4\lambda_{1S}^2 + 8\sqrt{M}\lambda_{1S}\lambda_{1T} + M(2+M)\lambda_{1T}^2 +16M\lambda_{2T}(\lambda_{1S}+\sqrt{M}\lambda_{1T}+M\lambda_{2T})).
\end{aligned}
\end{equation}
These beta function correspond to the large $N$ beta functions in \cite{Flodgren:2023lyl} with only a rescaling with respect to $M$. 
$M$ is a free parameter that is finite and affects the RG flow since it appears in the beta functions. 
The only fixed point for this system of equations is the origin.

This case has a closed dynamical system given by the quotient algebra $A/I^2$ which corresponds to the couplings space $(\lambda_{1S},\lambda_{1T})$. It has the idempotents (\ref{eq_idempotentsa0}) and non-trivial Peirce numbers (\ref{eq_PNa0}), both of which are real for $M\leq 2$. The values $M=0,1,2$ are special. For $M=0$ our multiplication table is singular, while for $M=1$ the idempotent $\textbf{c}^{0}=\textbf{c}^{-}$ and for $M=2$ the idempotent $\textbf{c}^{+}=\textbf{c}^{-}$.

\begin{figure}
     \centering
     \begin{subfigure}[b]{0.45\textwidth}
         \centering
         \includegraphics[width=\textwidth]{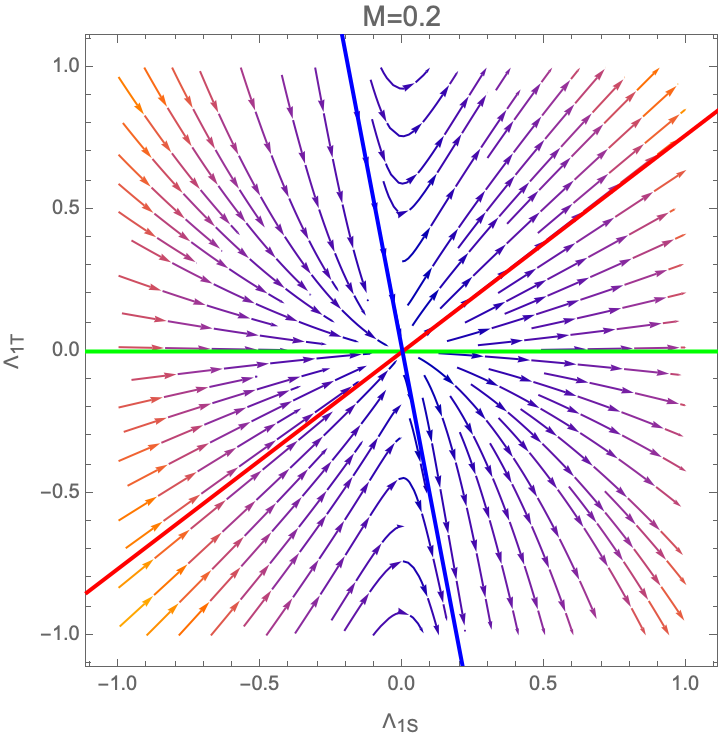}
         \caption{$M=0.2$}
         \label{fig:M1}
     \end{subfigure}
     \hfill
     \begin{subfigure}[b]{0.45\textwidth}
         \centering
         \includegraphics[width=\textwidth]{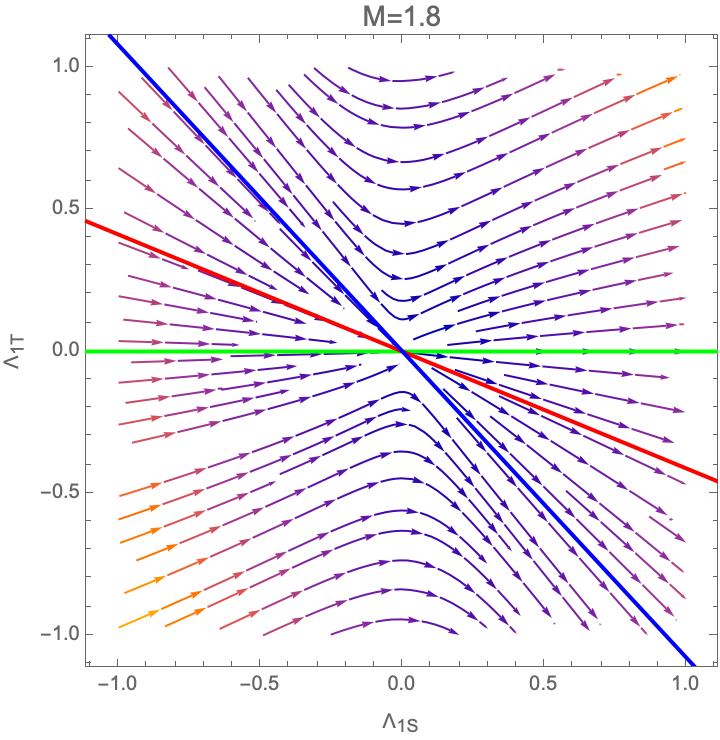}
         \caption{$M=1.8$}
         \label{fig:M2}
     \end{subfigure}
     \hfill
        \caption{RG flows for the case $a=0$. The figures show the 2D flows corresponding to the closed dynamical system $(\lambda_{1S},\lambda_{1T})$, for $M=0.2$ and $M=1.8$. The idempotents $\textbf{c}^0(M)$, $\textbf{c}^+(M)$ and $\textbf{c}^-(M)$ give the green, red and blue ray solutions respectively. As $M \rightarrow 1$ the red ray approaches the green since $\textbf{c}^0(1)=\textbf{c}^+(1)$. As $M \rightarrow 2$ the red and blue rays approach each other since $\textbf{c}^+(2)=\textbf{c}^-(2)$.
        }
        \label{fig:1S1T}
\end{figure}

\begin{figure}[!htb] 
\begin{center}
\includegraphics[scale=0.5]{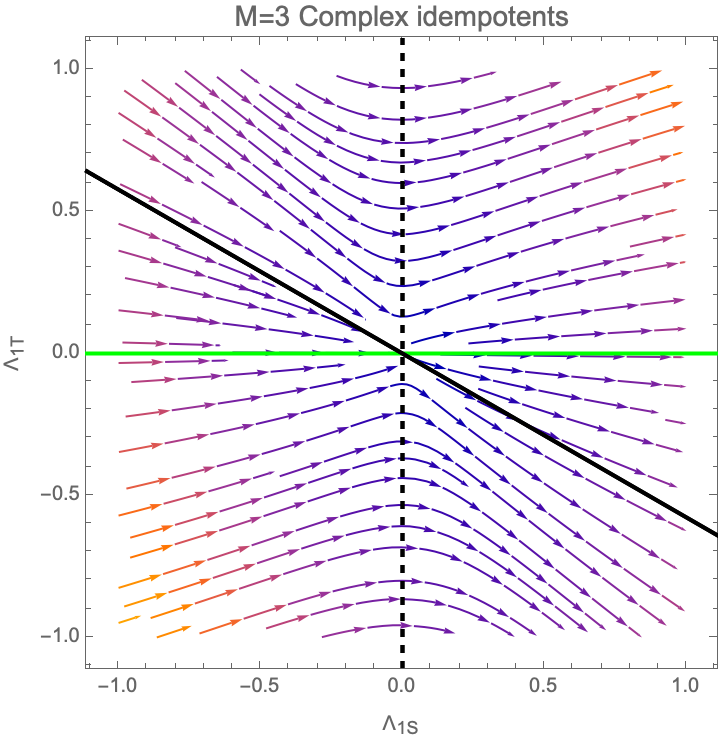}
\caption{RG flow for the case $a=0$. The figure shows the 2D flow corresponding to the $(\lambda_{1S},\lambda_{1T})$-space, for $M=3$. For $M>2$ the idempotents $\textbf{c}^+(M)$ and $\textbf{c}^-(M)$ are complex. The green ray corresponds to $\textbf{c}^0(M)$. The black line represents $\Re(\textbf{c}^+(M))=\Re(\textbf{c}^-(M))$ (which changes direction with $M$), and the dashed black line represents $\Im(\textbf{c}^+(M))=\Im(\textbf{c}^-(M))$ (which does not change direction with $M$).}
\label{Fig:1S1Ta0Comp}
\end{center}
\end{figure}

The RG flow for the quotient algebra $A/I^2$ is given by $\beta_{1S}$ and $\beta_{1T}$ and seen in figure \ref{fig:1S1T} for $M<2$. 
Figure \ref{Fig:1S1Ta0Comp} shows the RG flow for $M>2$, for which the idempotents $\textbf{c}^{\pm}$ are complex and have complex Peirce numbers. 

Complex Peirce numbers indicate a specific type of complex ray solution, but it is not obvious what the implications of complex idempotents and complex Peirce numbers are for the real solutions. Based on figure \ref{Fig:1S1Ta0Comp} (and figure \ref{fig:smallerv} for the case $a=2$) the imaginary part of a complex idempotent seems to indicate a line where the flow turns, while the real part might indicate a general direction in the flow. Figure \ref{fig:M2} shows a flow with general direction indicated by the red and blue rays which correspond to real idempotents. Figure \ref{Fig:1S1Ta0Comp} shows the real parts of the same idempotents (black line) which point in the same general flow direction.

As in the $a=2$ case an idempotent matches ray separatrices for $M<2$. When these rays solutions collide with other ray solutions at $M=2$ and the corresponding idempotents develop imaginary parts the real ray solutions disappear, and the separatrices which cannot disappear instead seem to curve.

\section{Conclusion}\label{conclusion}

We have seen that an algebraic description of the one-loop RG flows for a multiscalar theory characterises and organizes them. Typically, one may easily find a hierarchy of RG equations with some couplings inducing running in other couplings, but our analysis of the algebra of couplings and algebra ideals reveals more structure. There is a closed subset of couplings with decoupled RG equations of each ideal of the algebra. These ideals correspond to particular linear combinations of couplings. In the examples we have studied, large $N$ limits of the one-loop algebras in fact allow 
the identification of closed dynamical systems of couplings of lower dimensions than the original four dimensional system. There are closed three-dimensional systems and we have described dynamically closed two dimensional subspaces of couplings in detail.

For an adjoint multiscalar model with $SU(N)\times O(M)$ symmetry there are three distinct large $N$ limits that are identified by the algebra, the standard `t Hooft limit, a `multi-matrix' limit and an intermediate case. The algebras for these three cases have different free parameters and symmetries. It is satisfying that these large $N$ double scaling limits \eqref{eq_MN} gives examples of the three scalings discussed in \cite{Ferrari:2017jgw}: the standard matrix theory limit, the enhanced large $D$$(=M)$ limit of planar diagrams or ``new large $N$ limit'' of Ferrari \cite{Ferrari:2017ryl}, and a simple melonic tensorial limit, there considered for the case $D=M=N$, ie $a=1$. It is is reassuring that our simple consideration of scalings of a one-loop algebra captures the same limits which survived an advanced combinatorial analysis at all loop orders. Although it is not a proof, it supports the idea that the RG algebras are also relevant at higher orders.

Intriguingly, we have found an algebraic framework for the purely imaginary couplings proposed for `tetrahedral' couplings in trifundamental tensor models \cite{BenedettiGurauHarribey2019}. They appear to be associated with imaginary parts of idempotents of the extreme limit algebras with $a=0$ (standard `t Hooft) or $a=2$ (`multi-matrix') and to a real nilpotent of the intermediate ($0<a<2$) limit algebra. 
In the intermediate limit algebra the choice of a purely imaginary value for the coupling with `tetrahedral' properties leads to a line of non-trivial one-loop fixed points with real values for all other couplings. It is not yet clear if this pattern generalises to other large N models in other group representations or with other symmetry groups, but it shows that the algebra encodes interesting physics.\footnote{In the trifundamental case, as well as in our case, the one-loop line of fixed points is definitely a characteristic of perturbation theory, but it is not a physical property for finite couplings, as shown by two-loop results \cite{Benedetti:2020sye,BergesGurauPreis2023}.}

Not surprisingly, the RG flows we have found here in large $N$ limits of multiscalar theories are of similar form to the RG flows which have recently been proposed to represent asymptotically free scalar theories by Romatschke \cite{Romatschke2022}. It will be interesting to investigate this further.

Our work opens for several future directions. First, the full potential of the simplified properties of large $N$ RG flows will be realised only by systematically including gauge interactions, vector bosons, fermions and Yukawa couplings. Such a generalisation exists, since marginal couplings form a closed system under the RG flow, but the algebraic structure will be qualitatively different, because of the 3-point couplings, which do not appear in the marginal purely scalar case. 
A 3-point coupling receives a one-loop correction cubic in the 3-point coupling, suggesting that an algebraic formulation would require a 3-ary product rather than the binary $\diamond$ product, but it is not clear that this is the best way to proceed.

Second, taking corrections into account, at higher orders in $1/N$ and especially in the loop expansion, can be more challenging, but will be crucial for the future role of the algebra approach. Large $N$ corrections can be extracted directly from finite $N$ algebras, but the expansion should illuminate how simplifications at large $N$ emerge. In particular, it would be interesting to follow the fate of the decoupling of RG equations at finite $N$.

The most intriguing and potentially rewarding direction is the structure of higher loop corrections. An algebraic formulation of this structure can be very powerful, because it would sum even more different Feynman diagrams than are already summarised by the one-loop algebras. Again, the closure of the renormalisable and marginal couplings in perturbative RG theory reveals that there is structure connecting different loop orders.

\appendix

\section{Explicit form of tensor structure elements}\label{App0}
The four quartic invariant polynomials are
\begin{equation} \label{eq_polynomials}
\begin{split}
&\Tr \Phi_{\bar{a}} \Phi_{\bar{a}} \Phi_{\bar{b}} \Phi_{\bar{b}}, \Tr \Phi_{\bar{a}} \Phi_{\bar{b}} \Phi_{\bar{a}} \Phi_{\bar{b}}, \\ 
&\Tr \Phi_{\bar{a}} \Phi_{\bar{a}} \Tr \Phi_{\bar{b}} \Phi_{\bar{b}}, \Tr \Phi_{\bar{a}} \Phi_{\bar{b}} \Tr \Phi_{\bar{a}} \Phi_{\bar{b}}. 
\end{split}
\end{equation}
The tensor structure elements $e^{k}_{ABCD}$ are 
\begin{equation} \label{eq_4g}
\begin{aligned}
e^{1 s} \equiv & \delta_{\bar{a} \bar{b}} \delta_{\bar{c} \bar{d}}\left(\Tr\left(T_{\bar{A}} T_{\bar{B}} T_{\bar{C}} T_{\bar{D}}\right)+\Tr\left(T_{\bar{A}} T_{\bar{D}} T_{\bar{C}} T_{\bar{B}}\right)+\Tr\left(T_{\bar{A}} T_{\bar{B}} T_{\bar{D}} T_{\bar{C}}\right)+\Tr\left(T_{\bar{A}} T_{\bar{C}} T_{\bar{D}} T_{\bar{B}}\right)\right) \\
& +\delta_{\bar{a} \bar{c}} \delta_{\bar{b} \bar{d}}\left(\Tr\left(T_{\bar{A}} T_{\bar{B}} T_{\bar{D}} T_{\bar{C}}\right)+\Tr\left(T_{\bar{A}} T_{\bar{C}} T_{\bar{D}} T_{\bar{B}}\right)+\Tr\left(T_{\bar{A}} T_{\bar{C}} T_{\bar{B}} T_{\bar{D}}\right)+\Tr\left(T_{\bar{A}} T_{\bar{D}} T_{\bar{B}} T_{\bar{C}}\right)\right) \\
& +\delta_{\bar{a} \bar{d}} \delta_{\bar{b} \bar{c}}\left(\Tr\left(T_{\bar{A}} T_{\bar{C}} T_{\bar{B}} T_{\bar{D}}\right)+\Tr\left(T_{\bar{A}} T_{\bar{D}} T_{\bar{B}} T_{\bar{C}}\right)+\Tr\left(T_{\bar{A}} T_{\bar{B}} T_{\bar{C}} T_{\bar{D}}\right)+\Tr\left(T_{\bar{A}} T_{\bar{D}} T_{\bar{C}} T_{\bar{B}}\right)\right) \\
e^{1 t} \equiv & \delta_{\bar{a} \bar{b}} \delta_{\bar{c} \bar{d}}\left(\Tr\left(T_{\bar{A}} T_{\bar{C}} T_{\bar{B}} T_{\bar{D}}\right)+\Tr\left(T_{\bar{A}} T_{\bar{D}} T_{\bar{B}} T_{\bar{C}}\right)\right)+\delta_{\bar{a} \bar{c}} \delta_{\bar{b} \bar{d}}\left(\Tr\left(T_{\bar{A}} T_{\bar{B}} T_{\bar{C}} T_{\bar{D}}\right)+\Tr\left(T_{\bar{A}} T_{\bar{D}} T_{\bar{C}} T_{\bar{B}}\right)\right) \\
& +\delta_{\bar{a} \bar{d}} \delta_{\bar{b} \bar{c}}\left(\Tr\left(T_{\bar{A}} T_{\bar{B}} T_{\bar{D}} T_{\bar{C}}\right)+\Tr\left(T_{\bar{A}} T_{\bar{C}} T_{\bar{D}} T_{\bar{B}}\right)\right) \\
e^{2 s} \equiv & \delta_{\bar{a} \bar{b}} \delta_{\bar{c} \bar{d}} \delta_{\bar{A} \bar{B}} \delta_{\bar{C} \bar{D}}+\delta_{\bar{a} \bar{c}} \delta_{\bar{b} \bar{d}} \delta_{\bar{A} \bar{C}} \delta_{\bar{B} \bar{D}}+\delta_{\bar{a} \bar{d}} \delta_{\bar{b} \bar{c}} \delta_{\bar{A} \bar{D}} \delta_{\bar{B} \bar{C}} \\
e^{2 t} \equiv & \delta_{\bar{a} \bar{b}} \delta_{\bar{c} \bar{d}}\left(\delta_{\bar{A} \bar{C}} \delta_{\bar{B} \bar{D}}+\delta_{\bar{A} \bar{D}} \delta_{\bar{B} \bar{C}}\right)+\delta_{\bar{a} \bar{c}} \delta_{\bar{b} \bar{d}}\left(\delta_{\bar{A} \bar{B}} \delta_{\bar{C} \bar{D}}+\delta_{\bar{A} \bar{D}} \delta_{\bar{C} \bar{B}}\right)+\delta_{\bar{a} \bar{d}} \delta_{\bar{b} \bar{c}}\left(\delta_{\bar{A} \bar{B}} \delta_{\bar{D} \bar{C}}+\delta_{\bar{A} \bar{C}} \delta_{\bar{D} \bar{B}}\right),
\end{aligned}
\end{equation}
where the field indices have been suppressed on the left hand side and the normalisation is $\Tr(T_{\bar{A}}T_{\bar{B}})=\frac{1}{2}\delta_{\bar{A}\bar{B}}$. 
The polynomials (\ref{eq_polynomials}) are obtained by contracting the tensor structures (\ref{eq_4g}) with the fields $\phi_{A}\phi_{B}\phi_{C}\phi_{D}$. The exact relation is shown in (\ref{eq_poly1}).

\section{Algebra for finite $N$} \label{App1}
The products for the basis elements $\{e^k\}$, where $k=\{1s,1t,2s,2t\}$ and the large $N$ limit has not been taken, are
\begin{equation} \label{eq_Nprod1}%OBS: faktor 2 som saknades i (1t,1t)_(1t) är nu fixad.
\begin{split}
e^{1s}\diamond e^{1s} &= (\frac{-12-2M}{N} +\frac{N(M+3)}{2})e^{1s} -\frac{8}{N}e^{1t} + (\frac{(M+3)}{2} +\frac{M+4}{N^2} )e^{2s}+(\frac{1}{2}+\frac{2}{N^2})e^{2t} \\
 e^{1s}\diamond e^{1t} &= (\frac{-6-M}{N}+\frac{N}{2})e^{1s}+(-\frac{4}{N}+N)e^{1t}+(\frac{1}{2}+\frac{4+M}{2N^2})e^{2s}+(\frac{1}{2}+\frac{1}{N^2})e^{2t} \\
 e^{1t}\diamond e^{1t} &=  (\frac{N}{2}+\frac{-6-M}{2N})e^{1s} -\frac{2}{N}e^{1t} + (\frac{4+M}{4N^2})e^{2s} + (\frac{(M+2)}{8}+\frac{1}{2N^2})e^{2t},
\end{split}
\end{equation}
\begin{equation}\label{eq_Nprod2}
\begin{split}
 e^{2s}\diamond e^{2s} &= (M(N^2-1) +8)e^{2s}   \\
 e^{2s}\diamond e^{2t} &=   2(M+N^2)e^{2s}+6e^{2t}\\
 e^{2t}\diamond e^{2t} &= 12 e^{2s} + (2M+2N^2+6)e^{2t} ,
\end{split}
\end{equation}
and
\begin{equation}\label{eq_Nprod3}
\begin{split}
e^{1s}\diamond e^{2s} &= 6e^{1s} + (\frac{-2-M}{N}+(M+1)N)e^{2s}   \\
e^{1s}\diamond e^{2t} &= 2(M+3)e^{1s} + 8e^{1t}  +(-\frac{2}{N}+ 2N)e^{2s} + (-\frac{2}{N}+N)e^{2t} \\
e^{1t}\diamond e^{2s} &= 6e^{1t} +(\frac{-2-M}{2N} + N)e^{2s}   \\
e^{1t}\diamond e^{2t} &=  4e^{1s} + 2(M+1)e^{1t} -\frac{1}{N}e^{2s}+ (-\frac{1}{N}+N)e^{2t}. 
\end{split}
\end{equation}

\acknowledgments
We thank Vladimir Tkachev for alerting us about his suspicions regarding an earlier version of equations (\ref{eq_Nprod1}-\ref{eq_Nprod3}) for the finite $N$ algebra with consequences for the $a=2$ algebra in table \ref{Ta2}, and for sharing his knowledge about non-associative algebras. 
The work of B.S. is supported by the Swedish research council VR, contract DNR-2018-03803.

\bibliography{BibliographyManuscript}{}
\bibliographystyle{JHEP}

\end{document}